\documentclass[aps,prb,nobibnotes,nofootinbib,superscriptaddress,showkeys,showpacs]{revtex4}
\usepackage{amssymb,amsmath,bm}
\usepackage[english]{babel}
\usepackage[dvips]{graphicx}

\addto\captionsenglish{}
\addto\captionsrussian{}

\begin{document}

\title{Anomalous temperature and field behaviors of magnetization in cubic lattice frustrated ferromagnets}
\author{A.\,N.\,Ignatenko}
\email{ignatenko@imp.uran.ru}
\affiliation{Institute of Metal Physics, Kovalevskaya Str., 18, 620990, Ekaterinburg, Russia}
\author{A.\,A.\,Katanin}
\affiliation{Institute of Metal Physics, Kovalevskaya Str., 18, 620990, Ekaterinburg, Russia}
\affiliation{Ural Federal University, 620002, Ekaterinburg, Russia}
\author{V.\,Yu.\,Irkhin}
\affiliation{Institute of Metal Physics, Kovalevskaya Str., 18, 620990, Ekaterinburg, Russia}

\date{\today}

\begin{abstract}
Thermodynamic properties of cubic Heisenberg ferromagnets with competing exchange interactions are considered near the frustration point where the coefficient $D$ in the spin-wave spectrum $E_{\mathbf{k}}\sim D k^{2}$ vanishes. Within the Dyson-Maleev formalism it is found that at low temperatures thermal fluctuations stabilize ferromagnetism by increasing the value of $D$. For not too strong frustration this leads to an unusual "concave" shape of the temperature dependence of magnetization, which is in agreement with experimental data on the europium chalcogenides. Anomalous temperature behavior of magnetization is confirmed by Monte Carlo simulation. Strong field dependence of magnetization (paraprocess) at finite temperature is found near the frustration point.
\end{abstract}


\maketitle
\section{Introduction}
Recently the existence of a quantum disordered ground state in \emph{two-dimensional} magnets with predominantly \emph{ferromagnetic} interactions was demonstrated, see Refs. \cite{ShannonPRL06, Zhitomirsky_2010}. The formation of the corresponding quantum state, called a spin nematic, is associated with an instability with respect to the creation of bound states of spin waves. Since in the ferromagnetic ground state quantum fluctuations are completely absent, the physics of the "frustrated ferromagnets" differs substantially from the physics of antiferromagnets with competing exchange interactions.    

Magnetic frustrations also significantly influence the thermodynamic properties of three-dimensional systems, in particular the temperature dependence of long- and short-range order parameters (see e.g. Ref. \cite{JETPl_2008} for the fcc antiferromagnet). Anomalous temperature dependences of the magnetization are indeed observed experimentally in ferromagnetic materials, for example in the overdoped europium chalcogenides, see Ref. \cite{EuO_PRL2010}. Typically, these anomalies are explained by introducing into the Hamiltonian four-spin (biquadratic) interactions \cite{Nagaev}. In this paper we show that these anomalies can be naturally obtained in the framework of the standard Heisenberg model by taking into account the frustration consistently.

We consider the Heisenberg model to investigate thermal properties of three-dimensional frustrated ferromagnets on cubic latices. The frustration is caused by the exchange interactions in higher coordination spheres. Low-temperature properties of this model are determined by conventional magnons. The spin-wave spectrum $E_{0}(\mathbf{k})=S\left[J(0)-J(\mathbf{k})\right]$ (known exactly for a ferromagnet at zero temperature) is quadratic in the wave vector at $\mathbf{k}\to 0$, $E_{0}(\mathbf{k})\approx D_{0} k^{2}$, where $D_{0}=S\sum_{\mathbf{r}} J_{r}r^{2}/18$ is the spin-wave stiffness coefficient. Here $J(\mathbf{k})$ is the Fourier transform of the exchange interaction $J_{r}$, $r$ are the coordinates of the lattice sites. Provided that owing to the frustration $D_{0}$ vanish, there is a divergency in the equation for the magnetization 
\begin{equation}
\label{mag_eq}
M_{\text{SWT}} =S-\int\frac{d^{3} \mathbf{k}}{(2\pi)^3} \frac{1}{\exp{[E_{0}(\mathbf{k})/T]}-1}
\end{equation} 
in the spin-wave theory at finite temperature. Corrections to the spin-wave theory can be calculated systematically using the Dyson-Maleev representation for spin operators. 

\section{Results}
At low temperatures the spin-wave stiffness at finite temperature, $D(T)$, is determined by simple bubble diagram for self energy of magnons. Under normal conditions, i.e. in the absence of the frustration, the spin-wave stiffness decreases with increasing temperature and thus the magnetic order is destroyed. However, if the sufficient frustration is present, the situation change to directly opposite and ferromagnetic order at low temperatures is stabilized by thermal fluctuations \cite{JETPl_2013}. (similarly to the "order from disorder" effect, see, e.g., Ref. \cite{Villain1980}). In particular for $D_{0}=0$ we have obtained $D(T\to 0)=b T^{5/4}$, with $b>0$.

We have furthermore performed the calculations within the self-consistent spin-wave theory (SSWT), previously used for two- and three-dimensional systems \cite{IKK_PRB_USP}. SSWT includes all kinds of diagrams obtained from the simple bubble diagram by its recursive reinserting into internal lines. To improve the behavior of SSWT results at high temperatures, the contributions from pseudofermions are included, the chemical potential of bosons being introduced to fulfil the equation $M=0$ in the paramagnetic phase \cite{IKK_PRB_USP}. Fig. 1a shows an example of dependences $D(T)$ calculated in SSWT for $S=1/2$ Heisenberg model on the fcc cubic lattice. It can be seen that at low temperatures near the frustration point $D_{0}=0$ the dependences $D(T)$ are consistent with the above analytical result.
\begin{figure}[hpt]
\begin{center}
\includegraphics[width=8cm]{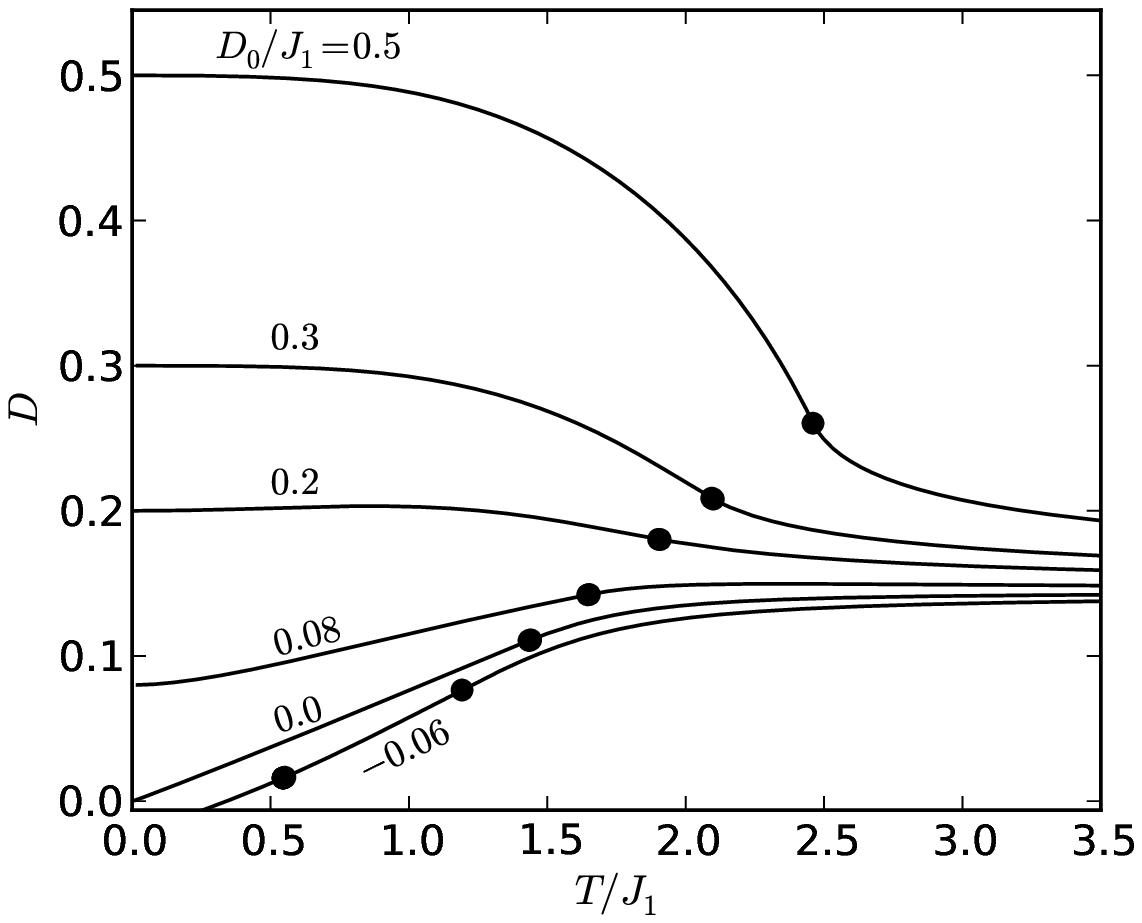}
\includegraphics[width=8cm]{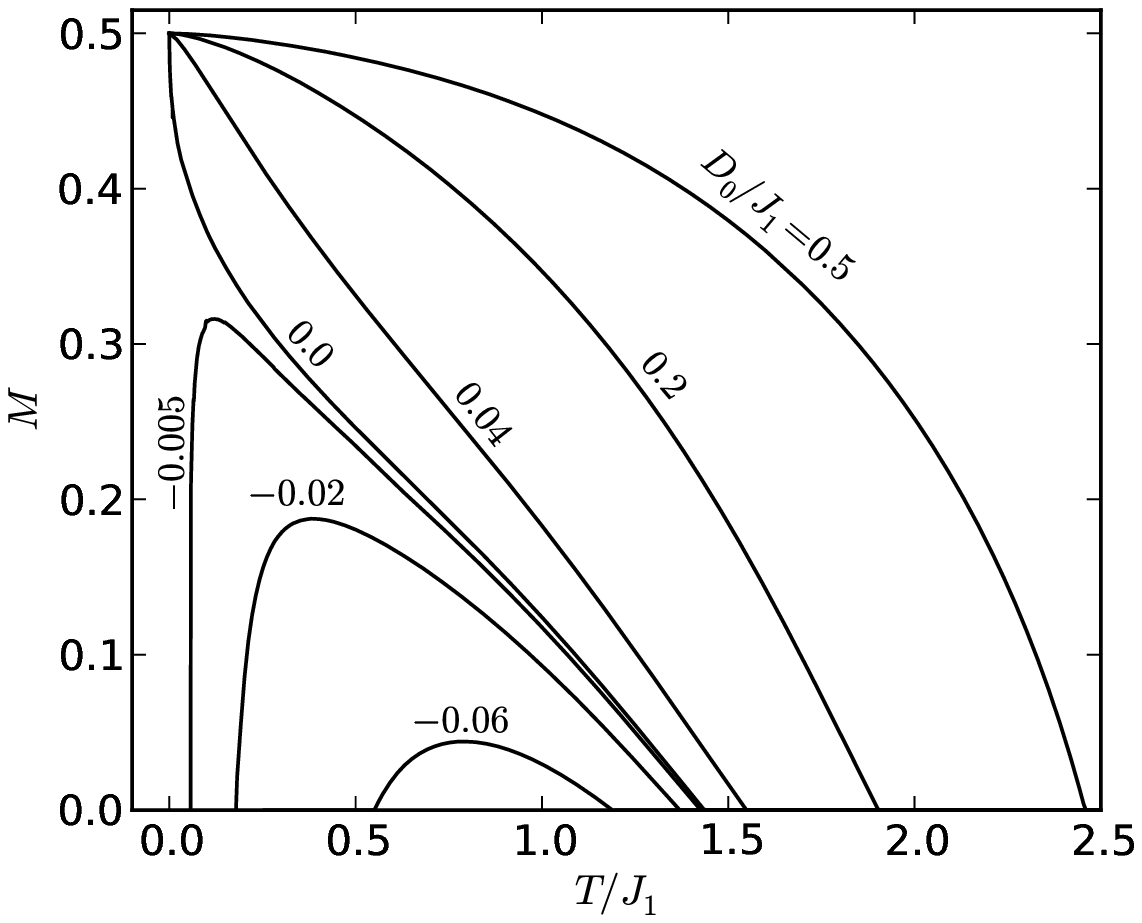}
\end{center}
\text{\hspace{0.27\textwidth} a \hspace{0.46\textwidth} b}
\caption{(a) Temperature dependence of the spin-wave stiffness calculated in SSWT for $S=1/2$ Heisenberg model on the fcc cubic lattice with the interaction between the first, second, and third neighbors. $J_{2}=-0.3 J_{1}$, and $J_{3}=\left(D_{0}/S-J_{1}-J_{2}\right)/6$ was chosen for a number of values of $D_{0}$ shown in the figure. The filled circles correspond to the Curie temperatures. (b) The dependence of magnetization on temperature in SSWT for the same model and parameters.}
\end{figure}

Note that the spectrum $E(\mathbf {k})$ in SSWT equation for the magnetization, is renormalized by temperature corrections. As a result, the magnetization does not diverge at $D_{0}=0$. However, the growth of the spin-wave stiffness with increasing temperature, $D(T\to 0)=b T^{5/4}$, results in this case in unusual behavior of the magnetization, $M\simeq S- uT^{3/8}$. Fig. 1b shows the temperature dependence of the magnetization for several values of $D_{0}$. It can be seen that for small non-negative $D_{0}$ the curves $M(T)$ at low temperatures acquire a concave form. Such anomalous dependences were observed experimentally in overdoped EuO \cite {EuO_PRL2010}.

The results for spin-wave stiffness and magnetization at negative $D_{0}$ (when the ferromagnetic ground state becomes unstable) are also shown in Figs. 1a and 1b. In this case the temperature corrections lead to the stabilization of the ferromagnetic ordered phase above some "lower" critical temperature. Therefore, the dependence of the Curie temperature $T_{c}$ on $D_{0}$ becomes two-valued at $D_{0}<0$. Note that classical systems can be studied in the same way as the quantum systems (for this it is necessary to replace $J_{ij}\to J_{ij}/S^{2}$ in all equations and then pass to the limit $S\to\infty$). All the results obtained in the quantum case do not change qualitatively in the classical limit (quantitative differences exist, in particular temperature exponents in equation for $D(T\to 0)$ will change). This is due to the weakness of the quantum effects in the ferromagnetic state. Below we will use this fact to identify the limits of applicability of SSWT by comparing SSWT with the results of computer simulation for classical systems.

When $D<0$ and ferromagnetism is unstable, another state, which is able to compete with the ferromagnetic state, should appear. In principle, this could be a spin liquid state. However, the most likely candidate is a spiral state. Note that for $D<0$ the spin-wave spectrum $E(\mathbf{k})\approx D k^{2}+\varkappa_{1}k^{4}+\varkappa_{2}(k_{x}^{4}+k_{y}^{4}+k_{z}^{4})$ has a minimum at $\mathbf{k}=\mathbf{Q}_{+}=(Q_{+},Q_{+},Q_{+})$, $Q_{+}=\left(-D/2[3\varkappa_{1}+\varkappa_{2}]\right)^{1/2}$, in the case $\varkappa_{2}>0$, and at $\mathbf{k}=\mathbf{Q}_{-}=(Q_{-},0,0)$, $Q_{-}=\left(-D/2[\varkappa_{1}+\varkappa_{2}]\right)^{1/2}$, in the case $\varkappa_{2}<0$. This could mean the formation of a spiral state with the wave vector $\mathbf {Q}_{+}$ or $\mathbf {Q}_{-}$. This state appears to be favourable in particular in the mean field approximation. Note that the ferromagnetic and paramagnetic states predicted in SSWT without account for spirals (which account is difficult) can be metastable. It is important that the continuous transition from ferromagnetic to spiral state is prohibited in SSWT because this transition would require $D(T)$ to vanish at the boundary between two phases. The latter is impossible due to the divergence in the magnetization for $D(T)\to 0$ and $T>0$.
\begin{figure}[bh]
\begin{center}
\includegraphics[width=8cm]{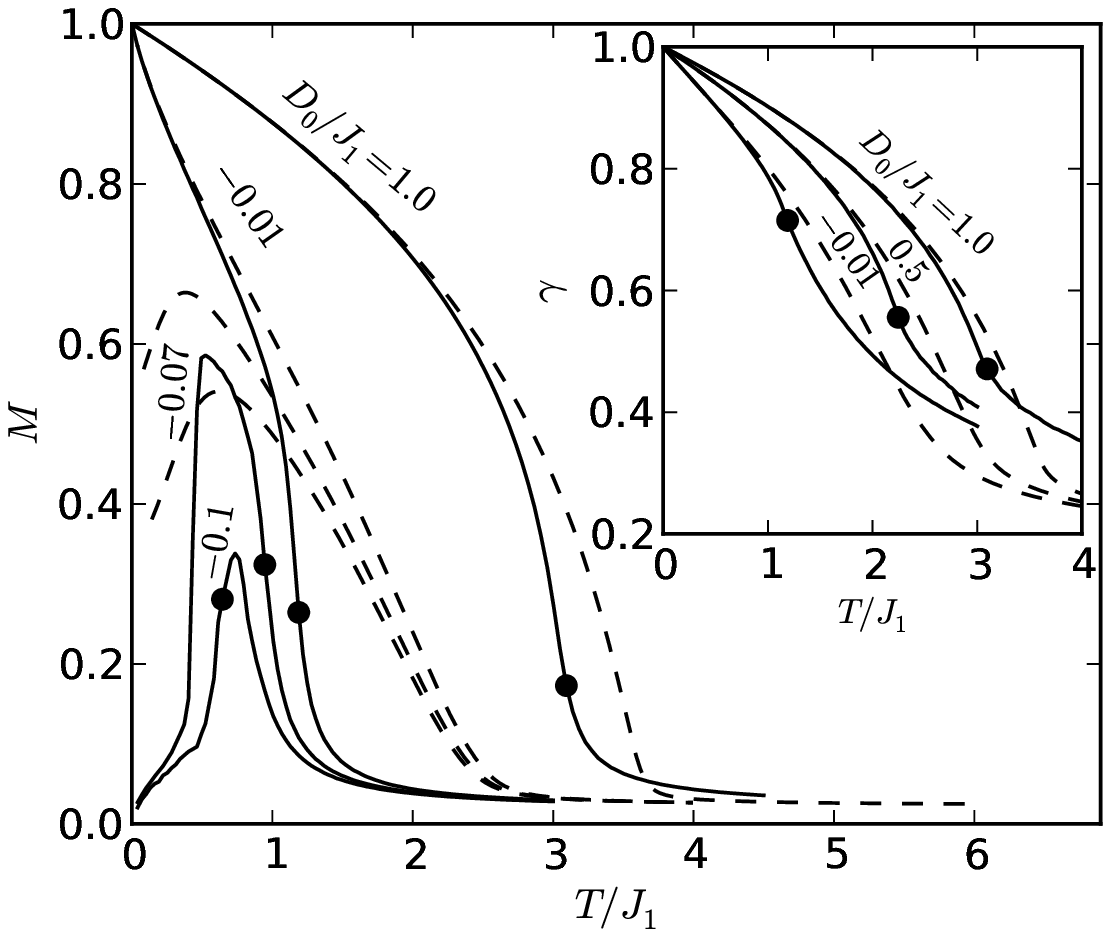}
\includegraphics[width=8cm]{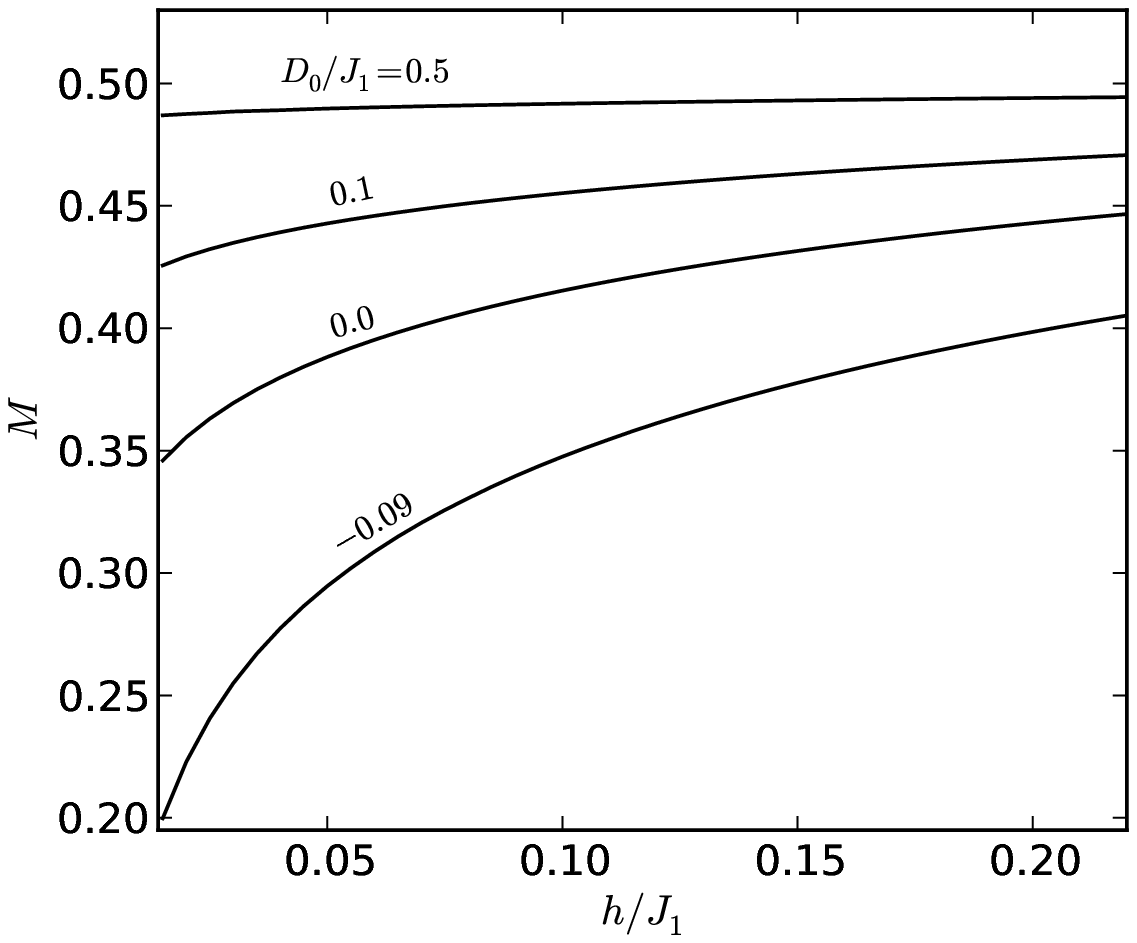}
\end{center}
\text{\hspace{0.27\textwidth} a \hspace{0.46\textwidth} b}
\caption{(a) Temperature dependence of the magnetization for the classical Heisenberg model on the finite $14\times 14\times 14$ fcc cubic lattice with the interaction of the first, second, and third neighbors, $J_{2}=-0.3 J_{1}$, $J_{3}=\left(D_{0}-J_{1}-J_{2}\right)/6$. The inset shows temperature dependences of the short-range order parameter. Solid lines are the results of Monte Carlo simulation. Dashed lines are the result of SSWT. The filled circles indicate the Curie temperatures determined by the maximums of the susceptibility. (b) Field dependence of the magnetization for the quantum Heisenberg model with $S=1/2$ on  fcc cubic lattice with the same exchange parameters as in left panel.}
\end{figure}

	To verify our results we have performed Monte Carlo simulation of the classical Heisenberg model on the finite fcc cubic lattices of extent $L=10, 14$ in each direction with. Fig. 2a shows that for $D_{0}\approx J_{1}$ SSWT gives a good description of the temperature dependence of the magnetization even at high temperatures, except for a narrow critical region near $T_{c}$. In the presence of frustration for $D_{\text{F}}(L)<D_{0}\ll J_{1}$ the ferromagnetic order in the ground state is stable, and SSWT still reliable at low temperatures, but becomes inapplicable at higher temperatures. The latter fact is manifested, in particular, in a significant overestimation of the Curie temperature. In the presence of strong frustration for $D_{0}<D_{\text{F}}(L)$ ferromagnetism in Monte Carlo simulation is observed only at intermediate temperatures, which is in a qualitative agreement with SSWT, but the quantitative discrepancies appear now at all temperatures.

	The inset in Fig. 2b shows the temperature dependence of the short-range order parameter $\gamma=\sqrt{\mathbf{S}_{i}\cdot\mathbf{S}_{j}}$ ($i$ and $j$ are nearest neighbors). In the presence of frustration the short range order persists up to $T_{c}$ and above, as one would expect. It is remarkable that even in the case of strong frustration and high temperatures SSWT describes the short-range order much better than the long-range one. This means that the basic defect of SSWT for strong frustration is associated with insufficient account for long-wavelength fluctuations. The existence of short-range order can be crucial for the energy gain of certain phases, which, in particular, is of interest for the description of the $\gamma$-$\alpha$ transition in iron \cite{JETPl_2008}.

We have also investigated the field dependence of magnetization in SSWT at finite temperature near the frustration point. The results are presented in Fig. 2b. One can see that for sufficiently strong frustration the dependence becomes considerably more strong which is somewhat analogous to the case of weak itinerant ferromagnets.

	Similar results can be obtained for all cubic lattices (simple cubic, bcc, and fcc) with different types of competing interactions. The presence of low-temperature regions with strong fluctuations on the phase diagram could favor the formation of a quantum disordered state in three-dimensional frustrated ferromagnets \cite{JETPl_2013}. To achieve advance in this direction, taking into account quantum effects in the spiral state near the boundary with ferromagnetic phase is of considerable interest.

\section{Acknowledgements}
	We are grateful to Yu.N.~Gornostyrev for valuable discussions. This work is partly supported by the Programs of fundamental research of RAS Physical Division "Quantum macrophysics and nonlinear dynamics", project No. 12-T-2-1001 (Ural Branch) and of RAS Presidium "Quantum mesoscopic and disordered structures", project No. 12-P-2-1041, and by ``Dynasty'' foundation. Monte Carlo simulation was performed using "Uran" cluster of IMM UB RAS.

\end{document}